\documentclass[twocolumn,twoside,slac_two]{revtex4}
\usepackage{graphicx}
\usepackage{fancyhdr}
\begin{document}

\title{TANAMI: Milliarcsecond Resolution Observations of Extragalactic Gamma-ray Sources}

%

\author{\mbox{Roopesh Ojha$^{1,2}$}, \mbox{Matthias Kadler$^{3,4,5,6}$}, \mbox{M. B\"ock$^{3,4}$}, \mbox{R. Booth$^{7}$}, \mbox{M. S. Dutka$^{8}$},\mbox{ P. G. Edwards$^{9}$}, \mbox{A. L. Fey$^{1}$}, \mbox{L. Fuhrmann$^{10}$}, \mbox{R. A. Gaume$^{1}$}, \mbox{H. Hase$^{11}$},\mbox{ S. Horiuchi$^{12}$}, \mbox{D. L. Jauncey$^{9}$}, \mbox{K. J. Johnston$^{1}$}, \mbox{U. Katz$^{4}$}, \mbox{M. Lister$^{13}$}, \mbox{J. E. J. Lovell$^{14}$}, \mbox{Cornelia M\"uller$^{3,4}$}, \mbox{C. Pl\"otz$^{15}$}, \mbox{J. F. H. Quick$^{7}$}, \mbox{E. Ros$^{10,16}$}, \mbox{G. B. Taylor$^{17}$}, \mbox{D. J. Thompson$^{18}$}, \mbox{S. J. Tingay$^{19}$}, \mbox{G. Tosti$^{20,21}$}, \mbox{A. K. Tzioumis$^{9}$}, \mbox{J. Wilms$^{3,4}$}, \mbox{J. A. Zensus$^{10}$}}
\affiliation{
$^{1}$\,United States Naval Observatory, 3450 Massachusetts Ave., NW, Washington DC 20392, USA\\
$^{2}$\,NVI, Inc., 7257D Hanover Parkway, Greenbelt, MD 20770, USA\\
$^{3}$\,Dr. Remeis-Sternwarte, Sternwartstr.~7, 96049 Bamberg, Germany\\
$^{4}$\,ECAP, Universit\"at Erlangen-N\"urnberg, Germany\\
$^{5}$\,CRESST/NASA Goddard Space Flight Center, Greenbelt, MD 20771, USA\\
$^{6}$\,USRA, 10211 Wincopin Circle, Suite 500 Columbia, MD 21044, USA\\
$^{7}$\,Hartebeesthoek Radio Astronomy Observatory, PO Box 443, Krugersdorp 1740, South Africa\\
$^{8}$\,The Catholic University of America, 620 Michigan Ave., N.E.,  Washington, DC 20064, USA\\
$^{9}$\,CSIRO Astronomy and Space Science, PO Box 76, Epping, NSW 1710 Australia\\
$^{10}$\,MPIfR, Auf dem H\"ugel 69, 53121 Bonn, Germany\\
$^{11}$\,BKG, Univ. de Concepcion, Casilla 4036, Correo 3, Chile\\
$^{12}$\,Canberra Deep Space Communication Complex, PO Box 1035, Tuggeranong, ACT 2901,  Australia\\
$^{13}$\,Dept. of Physics, Purdue University, 525 Northwestern Avenue, West Lafayette, IN 47907, USA\\
$^{14}$\,School of Mathematics \& Physics, Private Bag 37, Univ. of Tasmania, Hobart TAS 7001, Australia\\
$^{15}$\,BKG, Geodetic Observatory Wettzell, Sackenrieder Str. 25, 93444 Bad K\"otzting, Germany\\
$^{16}$\,Dept. d'Astronomia i Astrof\'{\i}sica, Universitat de Val\`encia, 46100 Burjassot, Val\`encia, Spain\\
$^{17}$\,Dept. of Physics and Astronomy, University of New Mexico, Albuquerque NM, 87131, USA\\
$^{18}$\,Astrophysics Science Division, NASA Goddard Space Flight Center, Greenbelt, MD 20771, USA\\
$^{19}$\,Curtin Institute of Radio Astronomy, Curtin University of Technology, Bentley, WA, 6102, Australia\\
$^{20}$\,Istituto Nazionale di Fisica Nucleare, Sezione di Perugia, 06123 Perugia, Italy\\
$^{21}$\,Dipartimento di Fisica, Universit\`a degli Studi di Perugia, 06123 Perugia, Italy\\
}

\begin{abstract}
The TANAMI (Tracking AGN with Austral Milliarcsecond Interferometry)
and associated programs provide comprehensive radio monitoring of
extragalactic gamma-ray sources south of declination $-30$
degrees. Joint quasi-simultaneous observations between the Fermi
Gamma-ray Space Telescope and ground based observatories allow us to
discriminate between competing theoretical blazar emission
models. High resolution VLBI observations are the only way to
spatially resolve the sub-parsec level emission regions where the
high-energy radiation originates. The gap from radio to gamma-ray
energies is spanned with near simultaneous data from the Swift
satellite and ground based optical observatories. We present early
results from the TANAMI program in the context of this panchromatic
suite of observations.

\end{abstract}

\maketitle

\thispagestyle{fancy}

\section{Introduction}

Near simultaneous observations across the electromagnetic spectrum are
essential to further our understanding of the physics of active
galactic nuclei (AGN). Milliarcsecond resolution Very Long Baseline
Interferometry (VLBI) observations make a unique contribution as they
are the only way to resolve the regions where the high-energy emission
originates. VLBI monitoring of AGN jets is the only way to directly
measure their relativistic motion and calculate intrinsic jet
parameters. VLBI observations let us probe the conditions under which
blazars and non-blazars emit $\gamma$-rays.

The TANAMI (\textbf{T}racking \textbf{A}ctive Galactic \textbf{N}uclei
with \textbf{A}ustral \textbf{M}illiarcsecond \textbf{I}nterferometry)
program (\cite{Ojha2008, Ojha2010}) provides parsec scale resolution
monitoring of extragalactic gamma-ray sources south of $-30$ degrees
declination at dual frequency (8.4 and 22\,GHz) by making VLBI
observations with the Australian Long Baseline Array (LBA\footnote{The
Long Baseline Array is part of the Australia Telescope which is funded
by the Commonwealth of Australia for operation as a National Facility
managed by CSIRO.}) and associated telescopes in Australia,
Antarctica, Chile and South Africa. Observations are made at intervals
of about two months. TANAMI observations are complemented by arcsecond
resolution monitoring across the radio spectrum with the Australia
Telescope Compact Array (PI: S. Tingay) and single-dish resolution
radio monitoring with the Hobart and Ceduna telescopes of the
University of Tasmania (PI: J. Lovell).

TANAMI began observations in November 2007 with an initial sample of
43 sources consisting of a radio selected flux-density limited
subsample and a $\gamma$-ray selected subsample of known and candidate
EGRET detections. This initial sample has since been expanded to
include new detections by the \textit{Fermi} Large Area Telescope
(LAT). Details of the sample and observations are presented in
\cite{Muller2009}. Here we outline some early results from the TANAMI 
program. 

\section{PKS\,1454-354}\label{sec:1454}

\begin{figure}
\includegraphics[width=70mm, clip]{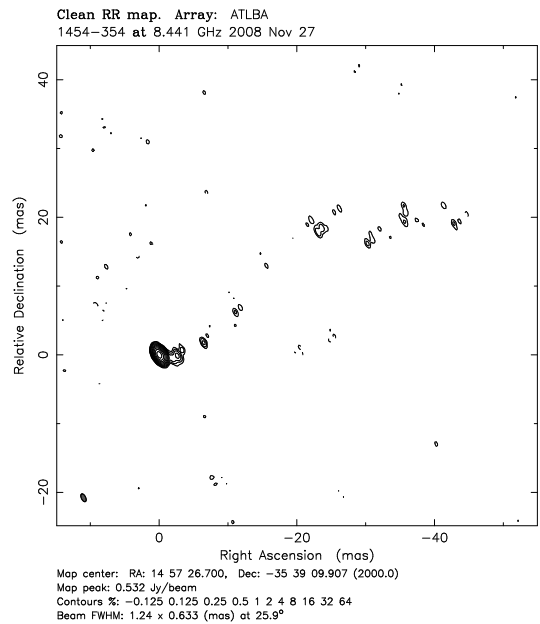}
\caption{Naturally weighted TANAMI image of PKS\,1454-354 at 8.4\,GHz. The image has an rms noise of $\sim 0.08$ mJy beam$^{-1}$. The hatched ellipse on the bottom left shows the restoring beam of $2.87 \times 0.75$ mas at $4^{\circ}$.}
\label{1454}
\end{figure}

On September 4th, 2008, a strong $\gamma$-ray flare was detected by
the Large Area Telescope (LAT) coming from the direction of the
flat-spectrum radio quasar PKS\,1454$-$354 ($z=1.424$). This quasar was
a possible counterpart of the unidentified EGRET source 3EG
J1500$-$3509 \cite{Hartman1999}. The flux rose on a timescale of hours
before dropping over the following two days. TANAMI images provided
the first high-resolution, high-sensitivity parsec-scale image of the
blazar jet for the first LAT AGN paper reporting this detection and
its multi-wavelength follow-up \cite{Abdo2009a}.

Figure~\ref{1454} shows a naturally weighted image of PKS\,1454$-$354
observed at 8.4\,GHz on Nov 10th, 2007. The image achieves a
resolution of $2.87 \times 0.75$ mas at an rms noise of $\sim 0.08$ mJy
beam$^{-1}$. It shows a compact core and a single-sided jet extending
to $\sim 50$ mas at a position angle of $\sim 300$, confirming the
activity of this source. Model fitting the core with an elliptical
Gaussian yields a brightness temperature limit of $T_{B} = 1.6 \times
10^{11}$K. The activity of this source continues to be monitored by
TANAMI at unprecedented resolution and dynamic range.

\section{Centaurus A}\label{sec:CenA}

\begin{figure}
\includegraphics[angle=-90, width=70mm, clip]{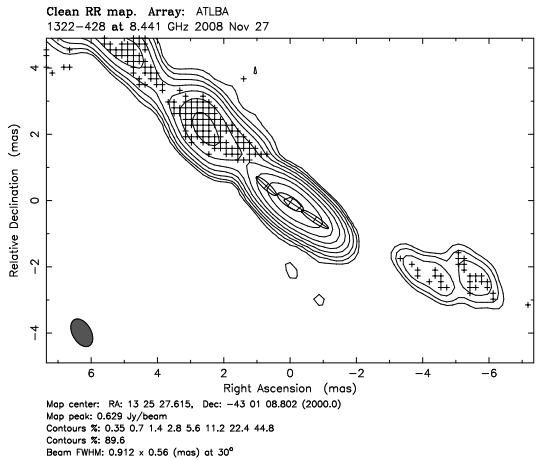}
\caption{The core region of Centaurus A at 8.4\,GHz. The best view yet of a possible $\gamma$-ray birthplace.}
\label{CenA_X}
\end{figure}

\begin{figure}
\includegraphics[angle=-90, width=70mm, clip]{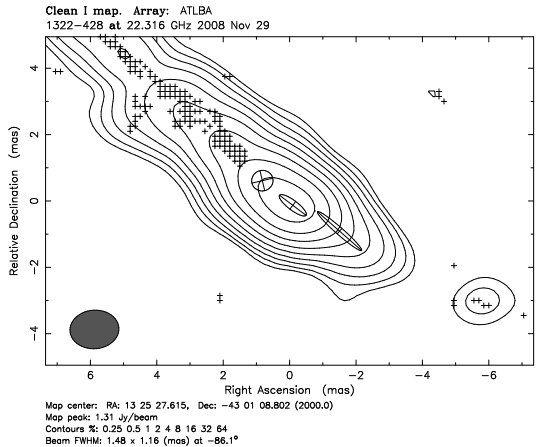}
\caption{The core region of Centaurus A at 22\,GHz. Note the close 
agreement with the core region at 8.4\,GHz shown in the previous 
figure.}
\label{CenA_K}
\end{figure}

At a distance of 3.4 Mpc \cite{Ferrarese2007}, Centaurus A is the
closest known radio galaxy making it possible to study it at subparsec
scales. Not surprisingly, it has been extensively observed at many
wavelengths and its basic VLBI-scale radio structure is well-known to
consist of a bright jet and a faint counter-jet at a viewing angle of
$50-80$ degrees \cite{Tingay1998}.

The TANAMI program has produced one of the highest resolution images
of an AGN jet ever made \cite{Kadler2010}. The uniformly weighted beam 
with a resolution
of $0.68 \times 0.41$\,mas happens to be comparable to the resolution
($0.8 \times 0.7$\,mas) of the previous highest resolution image of this
object \cite{Horiuchi2005}. Though these observations are separated by
about a decade the structures are remarkably similar which would
appear to be contrary to the previously reported apparent velocity of
$\sim 1.4$\,mas \cite{Tingay2001}. Velocity information obtained from
continuing multi-epoch TANAMI observations is being used to
investigate this.

Figure~\ref{CenA_X} and Figure~\ref{CenA_K} show the core region from 
naturally weighted images of 
Centaurus A at 8.4\,GHz and 22\,GHz respectively. In both cases the 
data have been self-calibrated using the CLEAN algorithm and the 
clean components in  the final model have been replaced by Gaussian 
model components in order to parametrize the core region. There is close
agreement between these two figures suggesting that the ``core'' has
the same size at both frequencies. Due to the presence of
trans-oceanic baselines to the TIGO (Chile) and O'Higgins (Antarctica)
telescopes only at 8.4\,GHz, the lower frequency image has the higher
resolution, resolving the core and providing the finest view of the
putative $\gamma$-ray production region seen so far. These simultaneous 
dual-frequency data are part of a set of simultaneous broadband data being 
used to model the behavior of Centaurus A \cite{Abdo2010}.


\section{First Epoch Results}

First epoch images at 8.4\,GHz of all 43 sources in the initial TANAMI
sample have been analyzed and the results are presented in
\cite{Ojha2010}. The images have high dynamic range and are often the 
best and sometimes the first images of an AGN at milliarcsecond scale 
resolution. Here we conclude this presentation with a few highlights 
from this analysis. Unless explicitly stated all discussions below 
are made with reference to the LAT 3-month list \cite{Abdo2009}.

\begin{itemize}
\item \textbf{Morphology and the \textit{Fermi} connection}\\
Figure~\ref{firstepoch} shows four sources which represent the four
morphological classes we have classified the TANAMI sample into. All
the images are made with naturally weighted, 8.4\,GHz data. The axes
are labeled in units of milliarcseconds. The restoring beam of each
image is shown as a hatched ellipse on the bottom left and, for the
three sources that have known redshifts, a bar indicating a linear
scale of 10\,pc or 1\,pc is shown on the bottom right. The
root-mean-square (rms) noise in the images is typically about
$\sim0.5$ mJy beam$^{-1}$. The lowest contour level is at 3 times the
root-mean-square noise level.

The classification scheme we have adopted is that of
\cite{Kellermann1998} which makes no assumptions about the physical
nature of the objects thus separating their description from their
interpretation. Clockwise from the top left, are examples of
single-sided (SS), compact (C), double-sided (DS) and Irregular (Irr)
morphology. The initial TANAMI sample has 33 SS sources and 5 DS
sources with just one example each of the other two morphological
types. Three sources do not have an optical identification. All of the
quasars and BL Lacertae objects in the sample have an SS morphology
while all 5 DS sources are galaxies. The lone C source is optically
unidentified while the only Irr source is a GPS galaxy 1718$-$649 
which is tentatively detected by LAT \cite{Boeck2009}.

\begin{figure*}[htbp]
\includegraphics[width=0.39\textwidth]{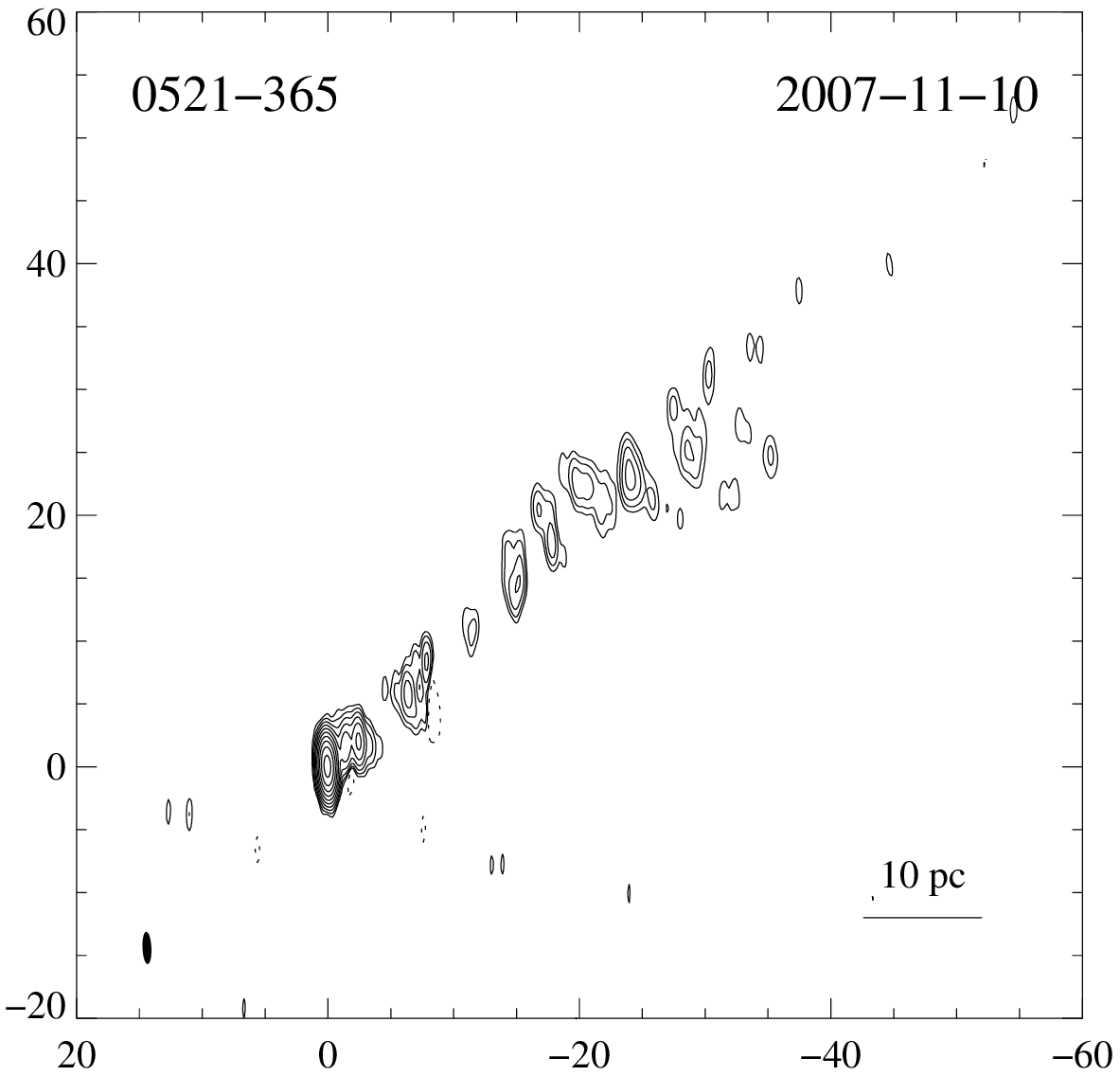}
\includegraphics[width=0.38\textwidth]{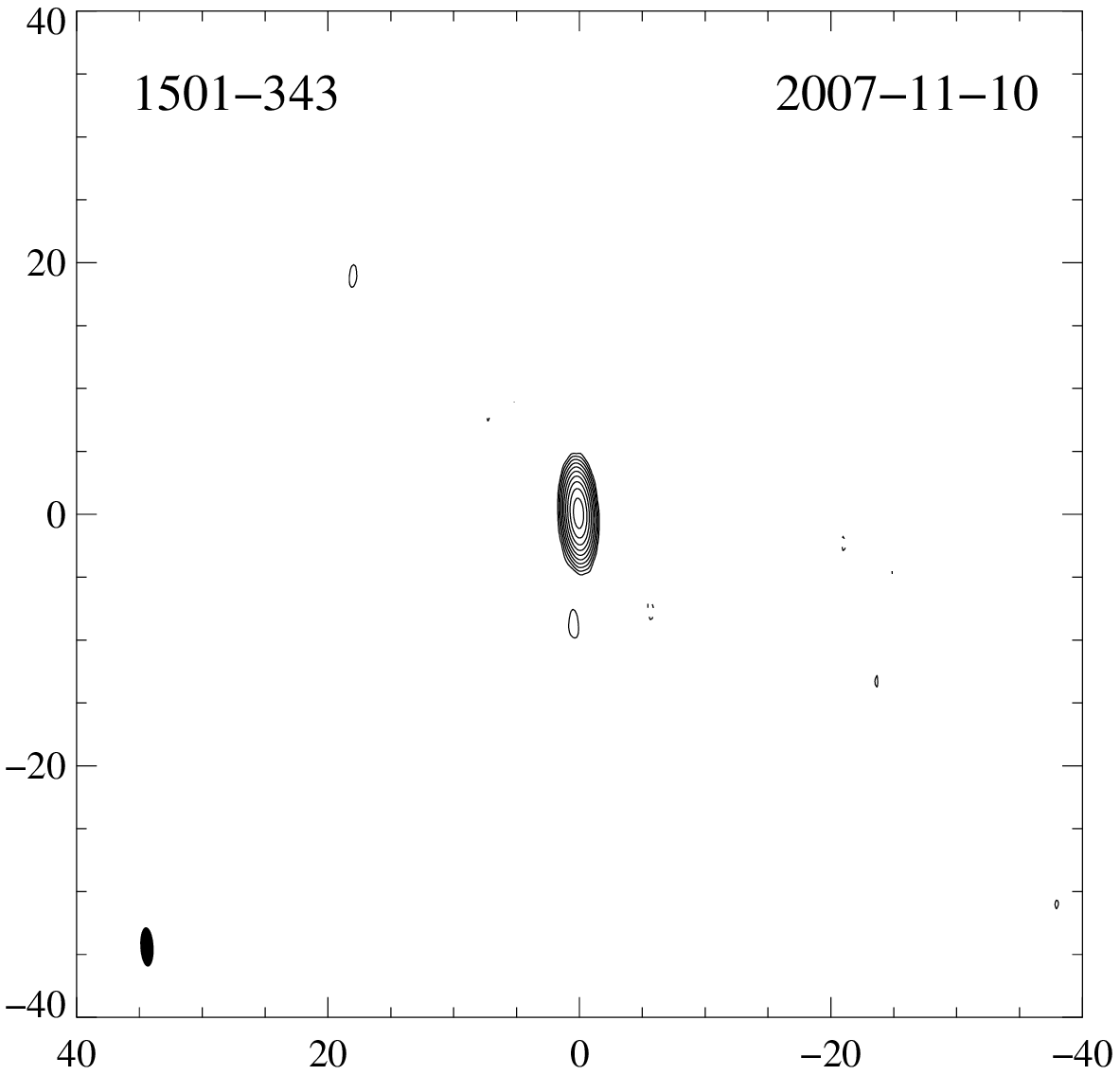}
\includegraphics[width=0.38\textwidth]{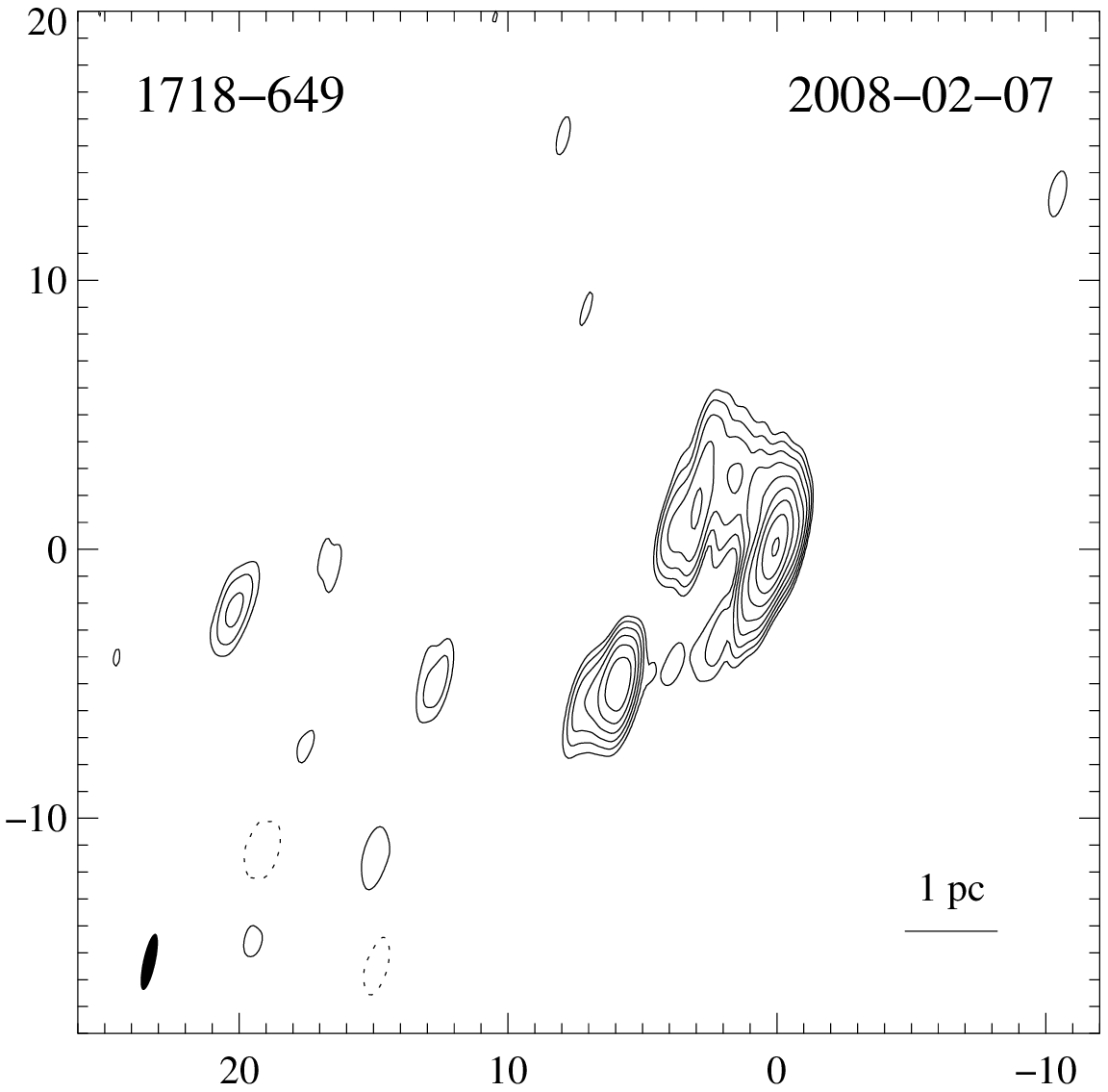}
\includegraphics[width=0.38\textwidth]{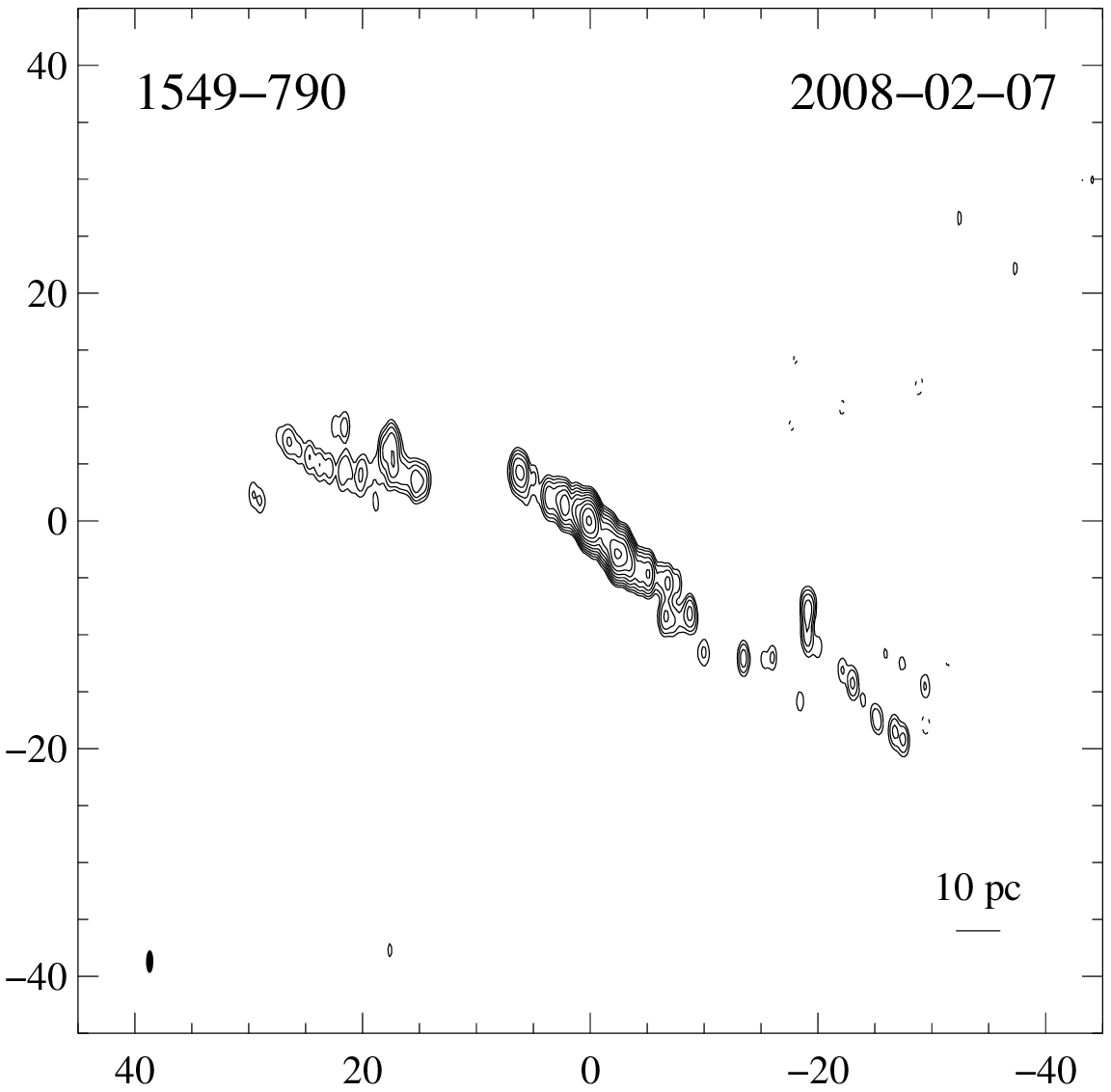}
\caption{TANAMI first epoch 8.4\,GHz images illustrating the different 
morphological types in the sample, see text for description. The 
scale of both axes of each image is in milliarcseconds and the 
restoring beam is shown as a hatched ellipse on the bottom left.}
\label{firstepoch}
\end{figure*}

In its first three months of operation \textit{Fermi} detected 12 of
43 TANAMI sources with $> 10 \sigma$ significance. All but one of
these sources are quasars and BLLacertae objects with an SS morphology
(the exception is the nearby radio galaxy Centaurus A). This accords
well with the canonical view that quasars and BL Lacertae objects have
an intrinsically symmetric twin-jet structure that is transformed by
differential Doppler boosting to the observed core-jet morphology.

\item \textbf{Are opening angles correlated with $\gamma$ luminosity?}\\
Given the expectation that the $\gamma$-ray emission from AGN is
beamed and thus orientation dependent, a link between $\gamma$-ray 
emission and the parsec scale morphology of AGN has been sought (e.g. 
\cite{Taylor2007}). We fit circular Gaussians to the visibility data 
and measured the angle at which the innermost jet component appears
relative to the position of the core i.e. the opening angle. Of the
LAT AGN Bright Sample (LBAS) sources $78\%$ have an opening angle $>
30$ degrees while only $27\%$ of non-LBAS sources do. In other words,
$\gamma$-ray bright jets are pointed closer to the line of sight than
$\gamma$-ray faint jets. This result should be treated with great
caution as the sample size for this analysis is currently small 
but \cite{Pushkarev2009} report similar results.

If confirmed the above result presents two possibilities: either the
LBAS jets have smaller Lorentz factors (since the width of the
relativistic beaming cone $\sim 1/\Gamma$) or LBAS jets are pointed
closer to the line of sight. The former scenario appears unlikely,
indeed the opposite effect is reported by \cite{Lister2009,
Kovalev2009}.

\item \textbf{Redshift distribution}\\
The redshift distribution of the quasars and BLLacs in the TANAMI
sample is similar to those for the LBAS and EGRET blazars. There does
not appear to be any significant difference between the radio- and
$\gamma$-ray selected subsamples.

\item \textbf{Brightness Temperature}\\
The core brightness temperature ($T_\mathrm{B}$) limit of all initial
TANAMI sources was calculated. The high end of the distribution of
calculated brightness temperatures is dominated by quasars and the low
end by BLLacertae objects and galaxies. Of the 43 sources in the
sample, 13 have a maximum $T_\mathrm{B}$ below the equipartition value
of $10^{11}$ K \cite{Readhead1994}, 29 below the inverse Compton limit
of $10^{12}$ K \cite{Kellermann1969}, putting about a third of the
values above this limit. Doppler boosting is the most likely reason
behind these high values though a variety of exotic mechanisms are
also possible. There is no significant difference in the brightness
temperature distribution of LBAS and non-LBAS sources.  Many of the
highest brightness temperature sources are not detected by LAT yet
which is counterintuitive since they are expected to have higher
Doppler factors.

\item \textbf{Luminosities}\\
The core and the total luminosity was calculated for all 38 initial
TANAMI sources that had published redshifts, assuming isotropic
emission. There is no significant difference in the distribution 
of luminosities of LBAS and non-LBAS sources. On the other hand, 
there is a clear relationship between luminosity and optical type 
with quasars dominating the high luminosity end of the distribution,
galaxies dominating the low luminosity end while the BL Lacertae 
objects fall in between,

None of the five most distant and most luminous sources have been
detected by Fermi in its first 3 months of operation. Intriguingly,
none of the nine most luminous jets (difference of total and core
luminosities) are detected. If this persists it would suggest a
completely unexpected anti-correlation between jet luminosity and
$\gamma$-brightness.

\end{itemize}

\section{Conclusion}
The TANAMI program is providing high quality, high resolution radio
monitoring for the southern third of the sky. Such data is a crucial
member of the suite of multiwavelength observations that is set to
revolutionize our understanding of the physics of AGN in the era of
\textit{Fermi}. 

Data from the TANAMI program is already a part of several individual
AGN studies and statistical analysis of the growing sample of TANAMI
sources is providing insight into the nature and origin of high energy
radiation from AGN. The availability of core and jet component spectra
as well as proper motions is going to greatly expand the questions
that TANAMI can help answer. 

\section{Acknowledgments} \label{Ack}
We are grateful to Dirk Behrend, Neil Gehrels, Julie McEnery, David
Murphy, and John Reynolds, who contributed in various ways to the
success of the TANAMI program so far. Furthermore, we thank the
\textit{Fermi}/LAT AGN group for the good collaboration.
This research has been partially funded by the Bundesministerium f\"ur
Wirtschaft und Technologie under Deutsches Zentrum f\"ur Luft- und
Raumfahrt grant number 50OR0808.  This research has made use of the
United States Naval Observatory (USNO) Radio Reference Frame Image
Database (RRFID).  This research has made use of data from the
NASA/IPAC Extragalactic Database (NED, operated by the Jet Propulsion
Laboratory, California Institute of Technology, under contract with
the National Aeronautics and Space Administration); and the SIMBAD
database (operated at CDS, Strasbourg, France). This research has made
use of NASA's Astrophysics Data System.


%
\end{document}